\documentclass{iopart}
\usepackage{epsfig}
\usepackage{subfigure}

\begin{document}
\newcommand{\figwidth}{0.95\columnwidth}
\newcommand{\ffigwidth}{0.4\columnwidth}
\newcommand{\figdir}{.}

\title[Sensitivity of protein rigidity analysis to small structural variations]{Sensitivity of protein rigidity analysis to small structural variations: a large-scale comparative analysis}

\author{Stephen A.\ Wells$^{1}$, J.\ Emilio Jimenez-Roldan$^{1,2}$, Rudolf A.\ R\"{o}mer$^{1}$}
\address{
 $^1$Department of Physics
  and Centre for Scientific Computing, University of Warwick, Coventry,
  CV4 7AL, United Kingdom
 $^2$Department of Systems Biology, University of Warwick, Coventry,
  CV4 7AL, United Kingdom\\
}

\date{$Revision: 1.38 $, compiled \today}

\begin{abstract}
$Revision: 1.38 $, compiled \today\\

Rigidity analysis using the ``pebble game'' can usefully be applied to protein crystal structures to obtain information on protein folding, assembly and the structure-function relationship. However, previous work using this technique has not made clear how sensitive rigidity analysis is to small structural variations. We present a comparative study in which rigidity analysis is applied to multiple structures, derived from different organisms and different conditions of crystallisation, for each of several different proteins. We find that rigidity analysis is best used as a comparative tool to highlight the effects of structural variation. Our use of multiple protein structures brings out a previously unnoticed peculiarity in the rigidity of trypsin.

\end{abstract}


\maketitle

\section{Introduction}
\label{sec-intro}

The ``pebble game'' \cite{JacT95} is an integer algorithm for rigidity analysis. By matching degrees of freedom against constraints, it can rapidly divide a network into rigid regions and floppy ``hinges'' with excess degrees of freedom. The algorithm is applicable to protein crystal structures if these are treated as molecular frameworks in which bond lengths and angles are constant but dihedral angles may vary; this application, and the program ``FIRST'' which implements the algorithm, have been described in the literature \cite{JacRKT01}. The rigid units in a protein structure may be as small as individual methyl groups or large enough to include entire protein domains containing multiple secondary-structure units (alpha helices and beta sheets). Rigidity analysis thus generates a natural multi-scale coarse graining of a protein structure \cite{GohT06}. The division of a structure into rigid units is referred to as a Rigid Cluster Decomposition (RCD).

This coarse graining has been used as the basis of simulation methods aiming to explore the large-amplitude flexible motion of proteins, first in the ROCK algorithm \cite{ThoLRJ01} and more recently in the ``FRODA'' geometric simulation algorithm \cite{WelMHT05}. FIRST/FRODA has been used to examine the inherent mobility of a protein crystal structure for comparison to NMR ensembles \cite{WelMHT05}, to examine possible mechanisms for the assembly of a protein complex \cite{JolWHT06}, to fit the structure of the bacterial chaperonin GroEL to low-resolution cryo-EM data \cite{JolWFT08}, and to examine the relation between protein flexibility and function in enzymes such as IDO \cite{MacNBC07} and myosin \cite{SunRAJ08}. The coarse-graining provided by the rigidity analysis dramatically reduces the computational cost of simulating flexible motion while retaining all-atom steric detail, making geometric simulation a promising complement to other methods such as molecular dynamics (MD).

Rigidity analysis itself (without motion simulation) has been used to study the formation of a virus capsid by assembly of protein subunits \cite{HesJT04} and, in particular, to examine the process of protein folding, considered as a transition from a floppy to a rigid state. The results of the rigidity analysis depend upon the set of constraints that are included in the network. While covalent bonds and hydrophobic tethers are always included in the analysis, the hydrogen bonds in the protein structure are assigned an energy using a potential based on the geometry of the bond \cite{DahGM97}, and only hydrogen bonds with energies below a user-defined ``cutoff'' are included in the constraint network. It is therefore possible to perform a ``rigidity dilution'' on a protein crystal structure, in which the ``cutoff'' is gradually reduced so that fewer and fewer hydrogen bonds are included in the analysis and the structure gradually loses rigidity.
 A study by Rader et al.\ \cite{RadHKT02} analysed a set of 26 different protein structures using rigidity dilution, drawing an analogy between the loss of rigidity during this dilution and the thermal denaturation of a protein. A similar study by Hespenheide et al.\ \cite{HesRTK02} made predictions for the ``folding core'' of several proteins based on the regions that retained rigidity longest during the dilution, obtaining a promising correlation with experimental data.

This variation of rigidity as different constraints are included, however, is in a sense the Achilles heel of the method if it is to be used as a coarse-graining approach and a basis for simulation. Firstly, it is unclear if there is a `correct' value of the hydrogen-bond energy cutoff, which can be used to obtain a physically realistic simulation of the flexible motion of a protein; the studies referenced above have used a wide variety of cutoff values on an {\it ad hoc} basis. Secondly, it is unclear how the RCD is affected by small variations in structure --- as, for example, if the same protein is crystallised under slightly different conditions, or if examples of a given protein from multiple different organisms are compared. Thus, we do not know if the RCD obtained for a given crystal structure and energy cutoff is in some sense `typical' of the protein, or if it is likely to vary dramatically compared to that of an apparently very similar structure and cutoff.

If the RCD of a protein is robust to small structural variations, this justifies the use of rigidity-based coarse graining for simulations of conformational change between structures \cite{JolWFT08} and structural variation due to flexible motion \cite{WelMHT05,MacNBC07,SunRAJ08}. However, if small changes in the structure cause dramatic changes to RCDs, then the RCD of one structure may not be representative. In this case, the rigidity analysis can best be used in a comparative mode; that is, to draw attention to parts of the structure where rigidity has changed, or to determine whether or not a given constraint (such as an interaction between two particular residues) is consistent with a given flexible motion. This was the approach taken by \cite{JolWHT06} when studying a major conformational change during assembly of a large protein complex. Previous studies on large numbers of proteins \cite{RadHKT02,HesRTK02} have considered single examples of multiple unrelated proteins, and hence do not provide the comparative information we need. A study by Mamonova et al.\ \cite{MamHST05} examined the variation in RCD of a protein structure in the course of a 10ns MD simulation, finding that the RCD varied quite dramatically for different snapshots of the MD trajectory.

An additional motivation for our study is to determine the typical pattern of rigidity loss during dilution. A recent study on rigidity percolation in glassy networks \cite{SarWHT07} found that the transition between rigid and floppy states could display either first-order or second-order behaviour. In the first case, the gradual introduction of constraints into the network led to a sharp transition from an entirely floppy state to one in which the entire system became rigid. In the second, rigidity initially developed in a percolating rigid cluster involving only a small proportion of the network, which then gradually increased in size as more constraints were introduced. By performing rigidity dilution on a very large number of protein structures and observing the pattern of rigidity loss, we should be able to determine whether the loss is typically sudden or gradual, and whether different types of protein display different behaviours.

 Our initial hypothesis was that proteins such as cytochrome C, whose function depends on retaining its structure to protect a contained heme group from exposure to solvent, would retain a large amount of rigidity over a wide range of cutoff values and then lose rigidity suddenly, while one whose function depends on conformational flexibility, e.g.\ an enzyme such as trypsin, would display a gradual variation in rigidity. As we shall see, the results we obtained were very different from our expectations.\footnote{``Give me fruitful error any time, bursting with its own corrections.'' --- Vilfredo Pareto.}

\section{Materials and Methods}
\label{sec-method}
\subsection{Protein selection}
\label{sec-select}

In this study we take a deliberately comparative approach by first choosing a set of globular proteins, and then, for each protein in the set, obtaining multiple crystal structures. We have particularly sought for (i) examples of the same protein from different organisms, {\it e.g.} cytochrome C proteins from multiple different eukaryotic mitochondria, and (ii) protein structures obtained under different conditions of crystallisation, e.g.\ in complex with different ligands or substrates.

The default treatment of hydrogen bonds and hydrophobic tethers in FIRST is based on the assumption that the protein exists in a polar solvent, e.g.\ cytoplasm, rather than being within a hydrophobic or amphiphilic environment as for membrane-bound proteins. Proteins in a membrane environment can still be handled but this requires hand-editing of the constraint network. In this study we have therefore confined ourselves to examining a set of non-membrane proteins. Our starting point was the set of proteins considered by Hespenheide et al.\ \cite{HesRTK02}. We then searched for proteins with similar structures, i.e.\ those listed under the same domain under SCOP classification and/or as homologous under CATH classification. We note that the RCSB Protein Data Bank provides these cross-references as derived information for each protein crystal structure \cite{BerWFG00}.

Rigidity analysis is best carried out on crystal structures with high resolution, so that we can have confidence in the accuracy of the atomic positions when constructing the hydrogen-bond geometries. We therefore eliminated NMR solution structures and concentrated on X-ray crystal structures with resolutions of better than $2.5$ \AA ngstroms. We also limited ourselves to wild-type structures rather than including engineered mutants.
To widen our data set we included several proteins not used in the earlier study. We included hemoglobin by analogy with myoglobin. The presence of a BPTI structure in the initial dataset let us to include a set of trypsins. In this paper we discuss results on 12 cytochrome C structures, 11 hemoglobin structures, 3 myoglobins, 6 alpha-lactalbumins and 18 trypsin structures. The full set of all data generated from our study (from over seventy crystal structures) is available as electronic supplementary information.

\subsection{Rigidity analysis method}
\label{sec-software}

We used the FIRST rigidity analysis software \cite{JacRKT01} to perform rigidity dilution (as described for example in \cite{RadHKT02}) on a wide selection of crystal structures obtained from the Protein Data Bank online database \cite{BerWFG00}. Our choice of protein crystal structures is described in section \ref{sec-select}. Each structure was processed as follows. From the crystal structure as recorded in the PDB, we extracted a single protein main chain, eliminating all crystal water molecules, but retaining important hetero groups such as the porphyrin/heme units of cytochrome C and hemoglobin. The ``PyMOL'' visualisation software \cite{Del____} proved very useful for this purpose. From multimeric crystal structures such as tetrameric hemoglobins, we extracted and used only a single protein main chain, reserving consideration of protein complexes to a future study which is now in preparation. We processed the resulting protein structure to add the hydrogens that are generally absent from X-ray crystal structures, using the ``REDUCE'' software \cite{WorLRR99} which also performs necessary flipping of side chains. After the addition of hydrogens we renumbered the atoms using ``PyMOL'' \cite{Del____} again to produce files usable as input to FIRST.

The program FIRST was then run over the resulting processed PDB file with appropriate command line options 
The energy of each potential hydrogen bond in the processed structure is calculated by FIRST using the Mayo potential \cite{DahGM97}, performing an initial rigidity analysis including all bonds with energies of $0$ kcal/mol or lower (option {\tt -E 0}). We also perform a dilution in which bonds are then removed in order of strength, gradually reducing the rigidity of the structure (option {\tt -dil 1}). 
The initial selection of proteins from the PDB was done by hand, but all subsequent stages were automated using bash scripts in a Linux environment with the jobs farmed out over a distributed cluster of workstations. This automation and the rapidity of the rigidity analysis software gave very high throughput for processing structures.

\begin{figure}
\begin{center}
\includegraphics[width=0.99\textwidth]{\figdir/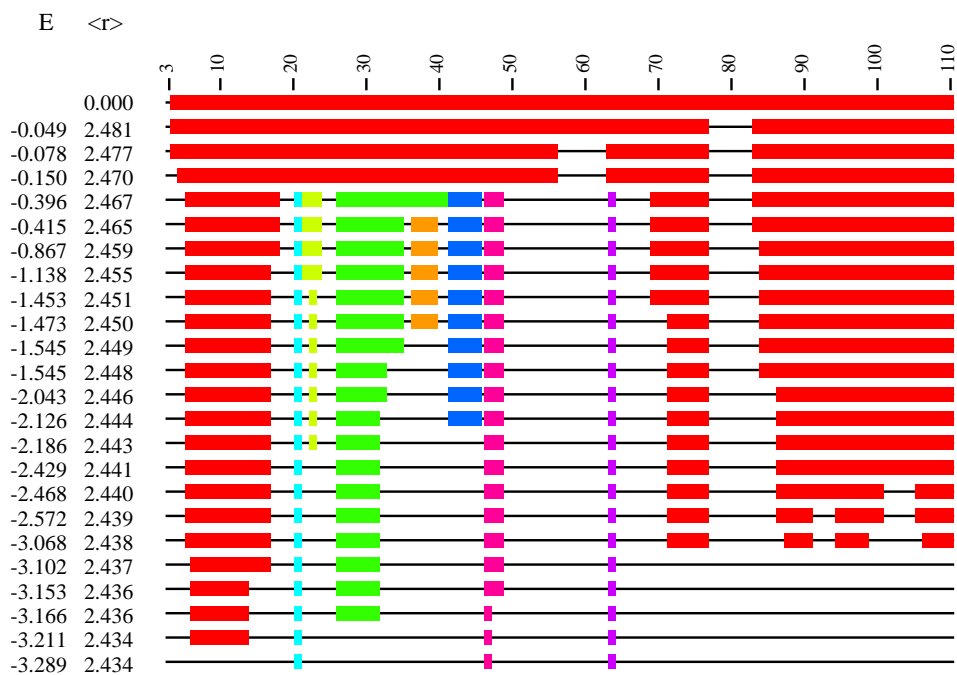}
\caption{Example of an RCD plot which is a result of the rigidity analysis. This plot is for the bacterial protein barnase, from the 1A2P.pdb structure. The labels have been lightly edited from the original FIRST output for clarity. The columns $E$ and $\langle r \rangle$ give the values of energy cutoff and network mean coordination, respectively, at which changes in the backbone rigidity occur. }
\label{stripy1}
\end{center}
\end{figure}

Running the rigidity dilution for a given protein produces an RCD plot (also called  `stripy plot') as illustrated in figure \ref{stripy1}. In this plot the horizontal axis represents the linear protein backbone. Flexible areas are shown as a thin black line while areas lying within a large rigid cluster are shown as thicker coloured blocks. The colour is used to show which residues belong to which rigid clusters; obviously the three-dimensional protein fold makes it possible for residues that are widely separated along the backbone to be spatially adjacent and form a single rigid cluster. The vertical axis on the RCD plot represents the dilution of constraints by progressively lowering the cutoff energy for inclusion of hydrogen bonds in the constraint network. Each time the rigid cluster analysis of the mainchain alpha-carbons changes as a result of the dilution, a new line is drawn on the RCD plot labelled with the energy cutoff and with the network mean coordination for the protein at that stage. We should stress that the RCD is always performed over the entire protein structure (mainchain and sidechain atoms) and a dilution is performed for every hydrogen bond removed from the set of constraints, typically several hundred bonds for a small globular protein. The RCD plot is then a summary concentrating on the rigid-cluster membership of the alpha-carbon atoms defining the protein backbone.

\subsection{Tracking rigidity}
\label{sec-tracking}

We can compare the results of the rigidity dilution on similar crystal structures in several ways. The most detailed form of comparison is to view the RCD plots for each structure side by side. This highlights the effect of structural variations; for example, a change in structure may lead what was previously a single rigid unit such as a helix to break up into two smaller rigid units connected by a flexible region. This is visible in the RCD plot as the division of a single coloured block into two. In this paper we will use this technique of detailed comparison to examine the rigidity of several mitochondrial cytochrome C structures.

This form of comparison, however, becomes unwieldy when comparing large numbers of structures. It is also not suitable for comparing results between different proteins whose structures are not closely related, as of course their rigidity-analysis RCD plots will not resemble each other at all. For larger-scale comparison we therefore adopt an approach inspired by recent studies on the rigidity properties of glassy networks. In this approach we extract a quantity measuring the degree of rigidity of the structure, and plot it as a function of either the hydrogen-bond energy cutoff, $E$, or the mean coordination of the network, $\langle r \rangle$. The latter quantity is the average number of covalently-bonded neighbours that each atom possesses in the network.
 In previous studies of glassy networks \cite{SarWHT07} the measure used was the fraction of atoms lying in the largest spanning rigid cluster in a model of a glassy network with periodic boundary conditions. This exact measure is inappropriate here as each protein structure is a finite network; we cannot define a spanning rigid cluster. Instead we extract the number of alpha-carbon atoms belonging to each of the largest rigid clusters, so that our measure of overall rigidity is the proportion of atoms that are found in large rigid units. The rigidity analysis sorts rigid clusters by size so that rigid cluster $1$ (RC1) is the largest.

The first measure that comes to mind is to track the number of number of alpha carbons lying within the single largest rigid cluster, $n$(RC1). However, this measure is clearly vulnerable to large variations. Consider the case of a large rigid cluster which divides, due to the removal of a constraint. At one extreme, a small portion of the cluster may become flexible leaving the cluster size only slightly reduced. At the other extreme, the cluster may divide into two clusters of roughly equal size linked by a flexible hinge; in this case the size of the largest rigid cluster is roughly halved. It would be hard, however, to argue that one case is a greater loss of protein rigidity than the other. We have therefore chosen a less sensitive measure; the number of alpha-carbon atoms found within the first {\em five} rigid clusters, denoted as $n$(RC1-5). Examination of the data files confirms that our results would not change significantly if we were to choose the first four or siz rigid clusters instead; the use of the first five largest rigid clusters appears, for these small globular proteins, to capture all clusters containing any significant number of backbone alpha-carbon atoms.

This measure is not directly reported by the FIRST software during a dilution. We obtained our data by first running the dilution over all structures as described in section \ref{sec-method}; then, for each structure, we extracted the values of the energy cutoff at which changes in the backbone rigidity were observed. We then again run rigidity analyses at each of these cutoff values and extract our rigidity measure.

\begin{figure}
\begin{center}
\includegraphics[angle=270,width=0.49\textwidth]{\figdir/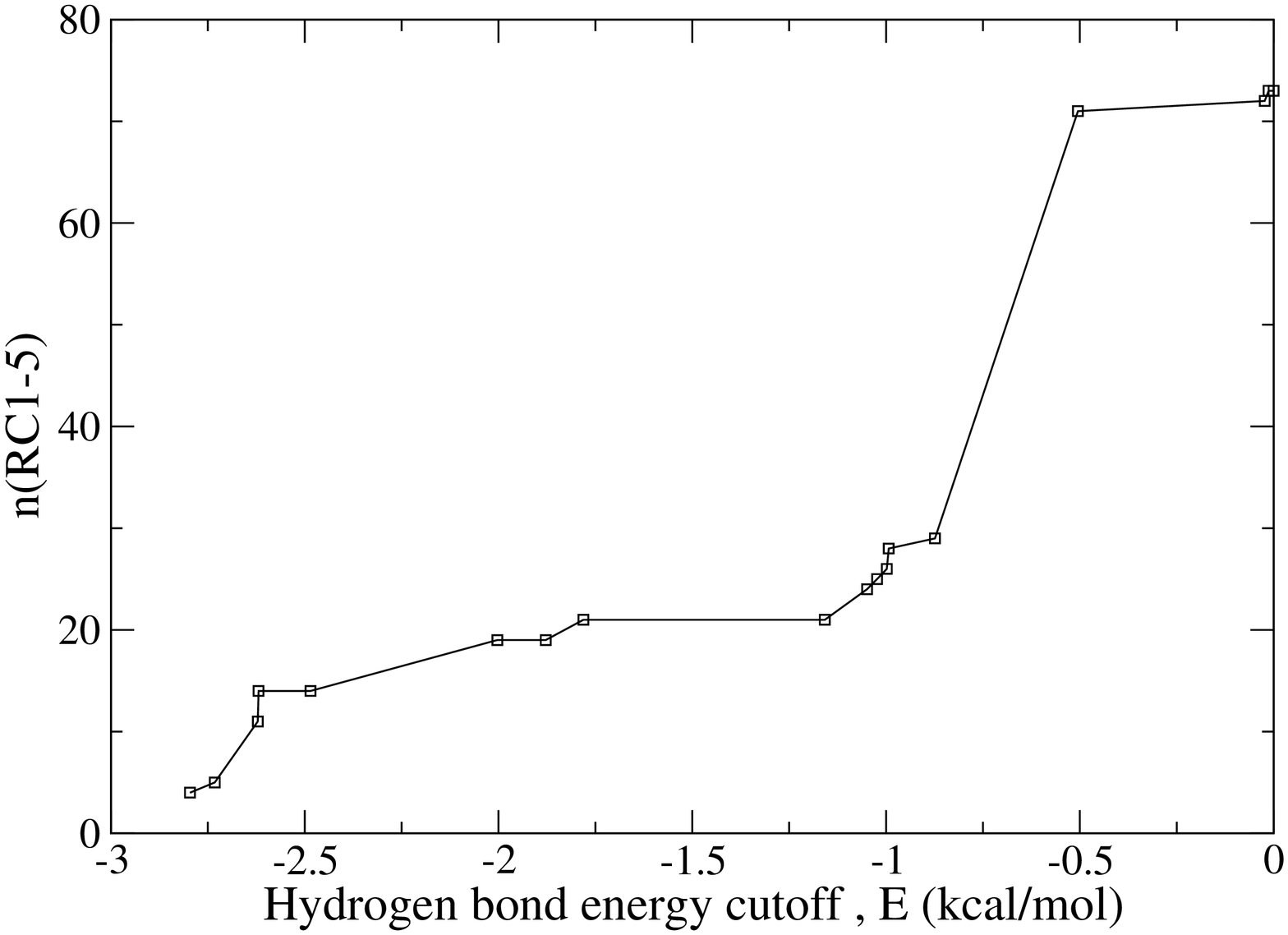}
\includegraphics[angle=270,width=0.49\textwidth]{\figdir/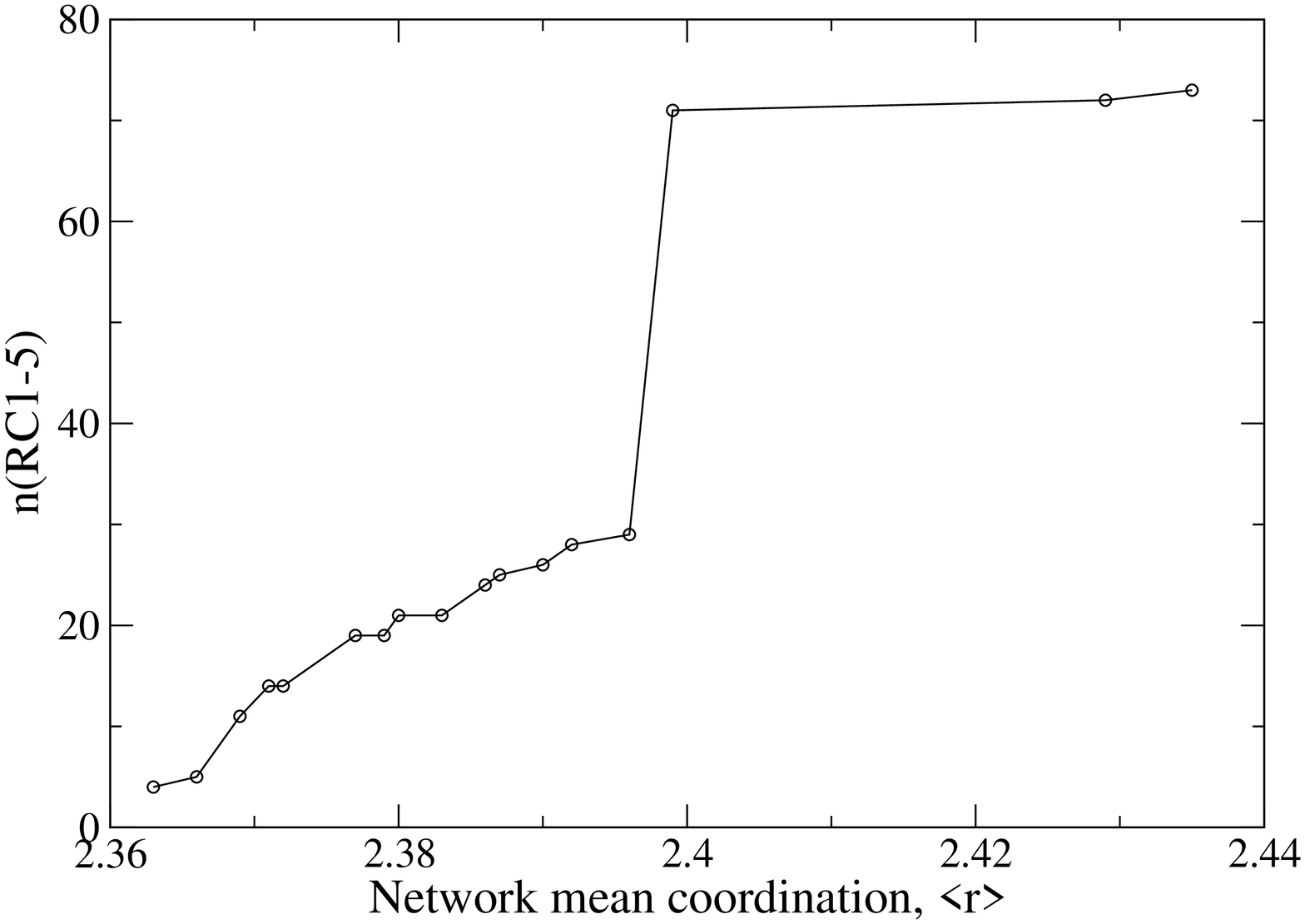}
\caption{Rigidity $n$(RC1-5) versus hydrogen-bond energy cutoff $E$ and network mean coordination $\langle r \rangle$ for human ubiquitin, 1UBI. Here and in all following such graphs, lines are guides to the eyes only.}
\label{fig-example-en}
\label{fig-example-rn}
\end{center}
\end{figure}

We can plot rigidity as a function either of the hydrogen-bond energy cutoff $E$, or as a function of the network mean coordination $\langle r \rangle$ (cp.\ figure \ref{fig-example-en}). In the latter case we see results reminiscent of glassy networks \cite{SarWHT07}, with the structure being largely rigid when the network mean coordination exceeds $2.4$ and becoming more flexible at lower mean coordinations. In the case of proteins, however, finite-size effects and open boundary conditions mean that flexibility can persist at mean coordinations above $2.4$. In this study we will concentrate on the behaviour of rigidity as a function of energy cutoff, the variable which is directly under the user's control when using the FIRST software.

\section{Results and discussion}
\label{sec-results}

We begin with a detailed comparison of cytochrome C structures in section \ref{sec-cytC}. Examination of four horse mitochondrial cytochrome C structures (1HRC, 1WEJ, 1CRC and 1U75) obtained under different conditions of crystallisation, and then of four tuna mitochondrial cytochrome C structures (1I55, 1I54, 5CYT and 1LFM) with different heme-group metal ions, gives us a sense of how much variation in rigidity can result from small variations in crystal structure. We then compare a wide set of cytochrome C structures from eukaryotic mitochondria (1I55, 1CYC, 1YCC, 2YCC and 1CCR) and a bacterium (1A7V) to determine if there is any correlation between the phylogenetic similarity of proteins and their rigidity behaviour.

In section \ref{sec-multi} we will examine a set of hemoglobin structures including four human structure in different oxidation states (1A3N deoxy, 2DN1 oxy, 2DN2 deoxy and 2DN3 carbonmonoxy) and a selection of structure from other eukaryotes including goose (1A4F), bovine (1G09), worm (1KR7), clam (1MOH), alga (1DLY), protozoa (1DLW) and rice (1D8U). We will also consider a set of myoglobins from sea turtle (1LHS), sperm whale (1HJT) and horse (1DWR) and a set of alpha-lactalbumin structure derived from several mammals: human (1HML, 1A4V), baboon (1ALC), goat (1HFY), bovine (1F6R) and guinea-pig (1HFX).

Finally, in section \ref{sec-trypsin}, we will consider a wide-ranging set of trypsin structures from humans (1H4W, 1TRN), rat (1BRA, 1BRB, 1BRC, 3TGI), pig (1AVW, 1AVX, 1LDT), salmon (1A0J, 1BZX) and bovine (in complex with various inhibitors --- 1AQ7, 1AUJ, 1AZ8 --- and small molecules --- 1K1I, 1K1J, 1K1M, 1K1N, 1K1O, 1K1P). The behaviour of trypsin structures during rigidity dilution appears distinctively different from the other proteins we have considered.

\subsection{Cytochrome C}
\label{sec-cytC}

\begin{figure}
\begin{center}
  \subfigure[1HRC: uncomplexed]{\includegraphics[scale=0.45]{\figdir/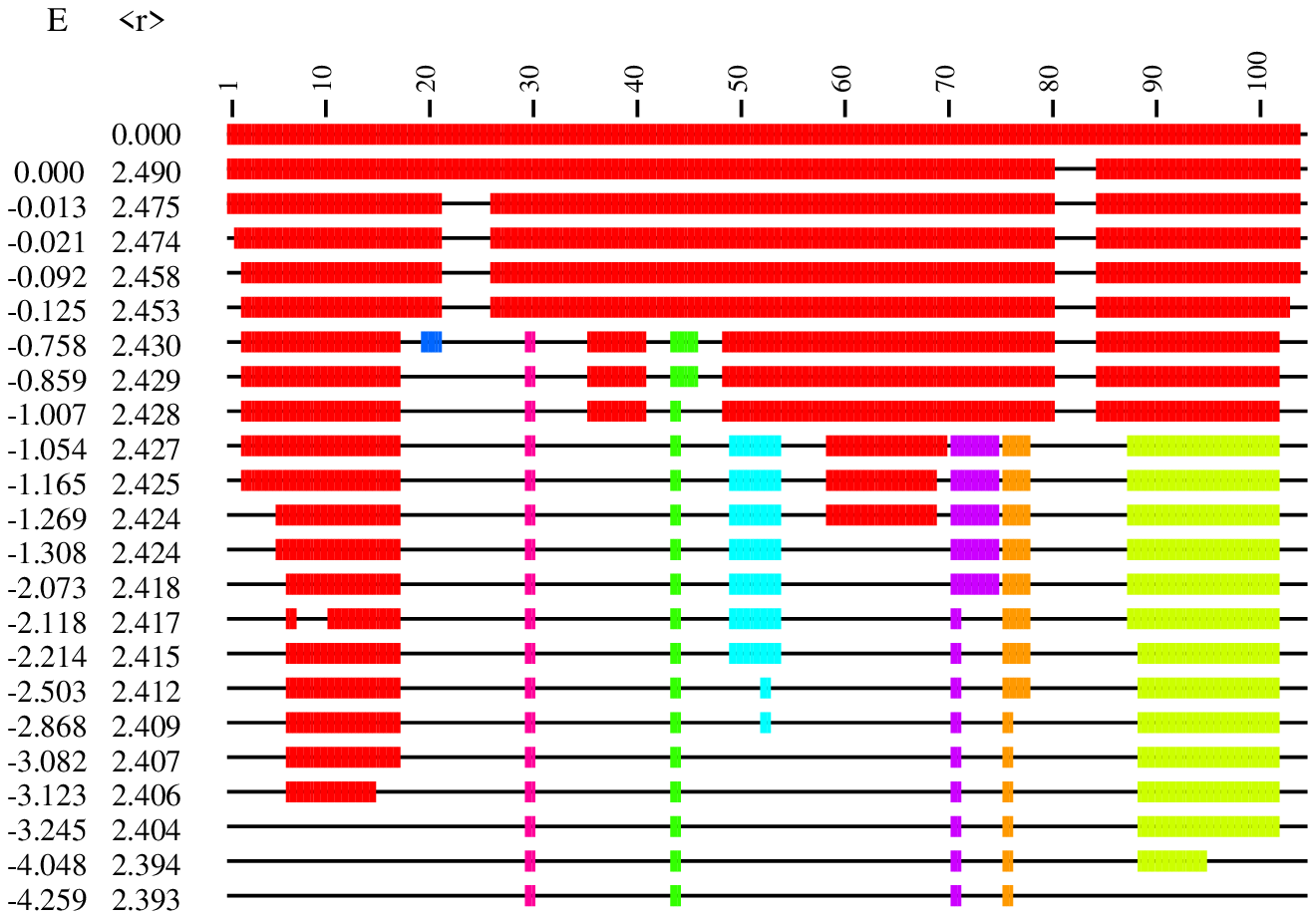} }
  \subfigure[1WEJ: antibody complex]{\includegraphics[scale=0.45]{\figdir/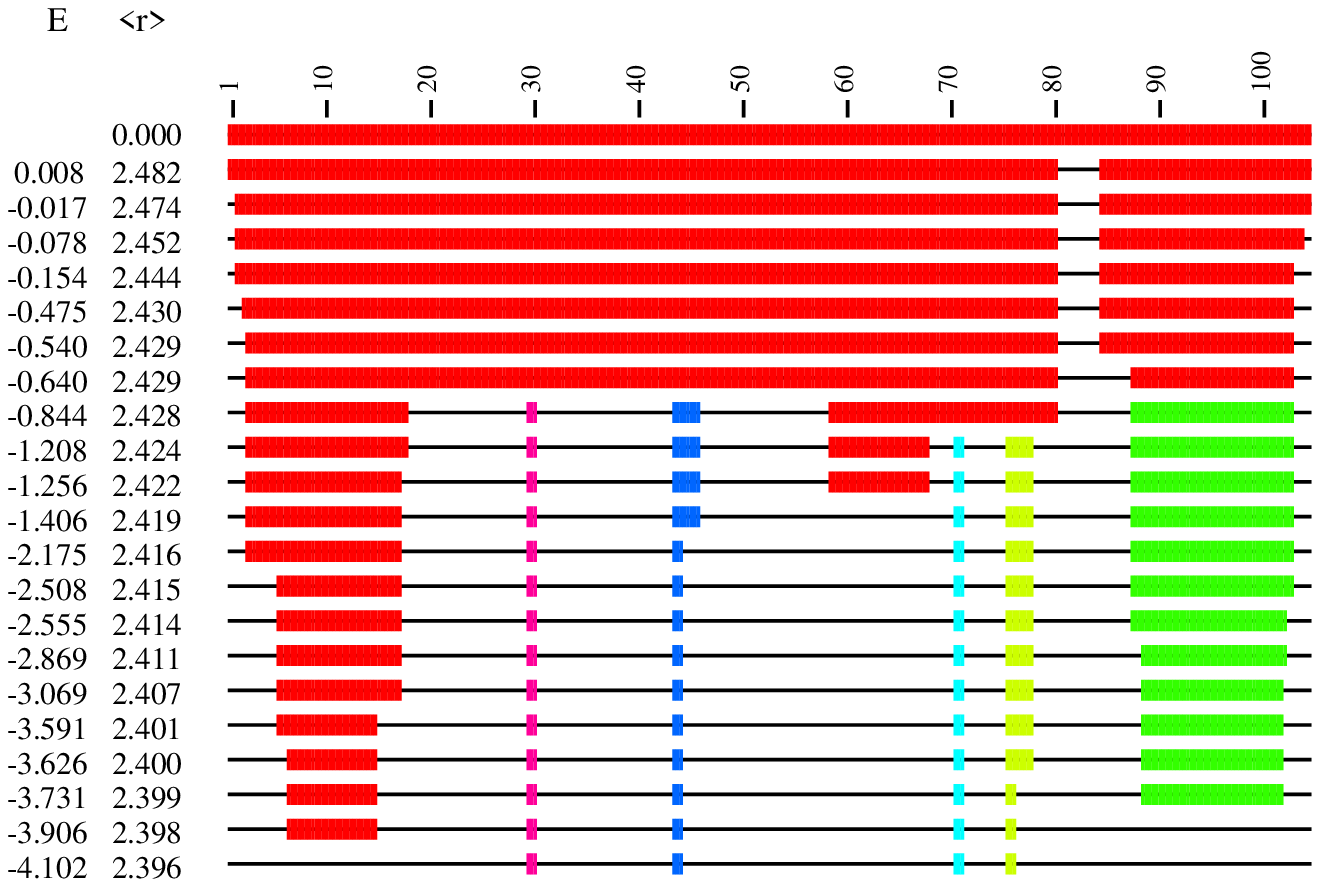} }
  \subfigure[1U75: peroxidase complex]{\includegraphics[scale=0.45]{\figdir/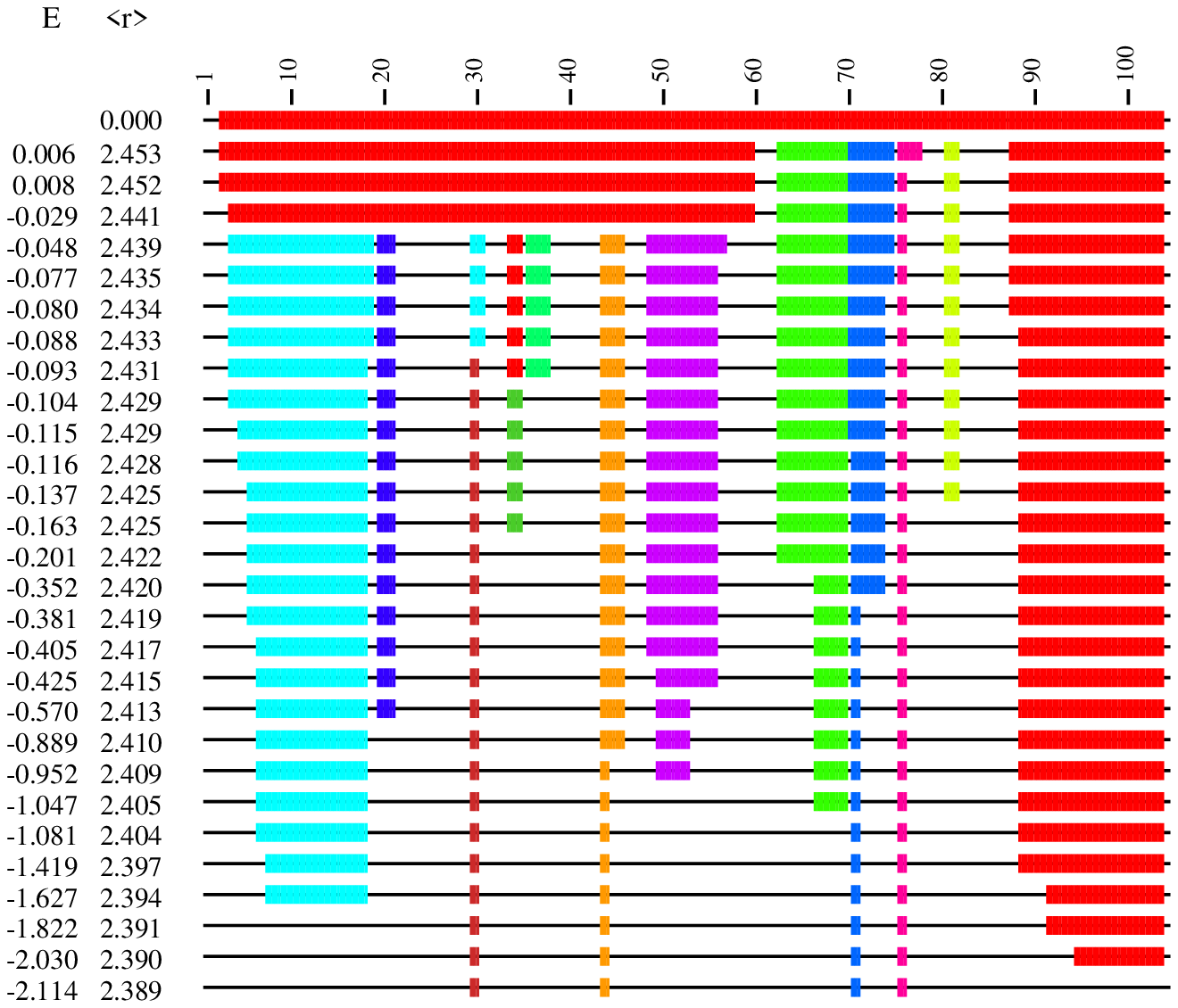} }
  \subfigure[1CRC: low ionic strength]{\includegraphics[scale=0.45]{\figdir/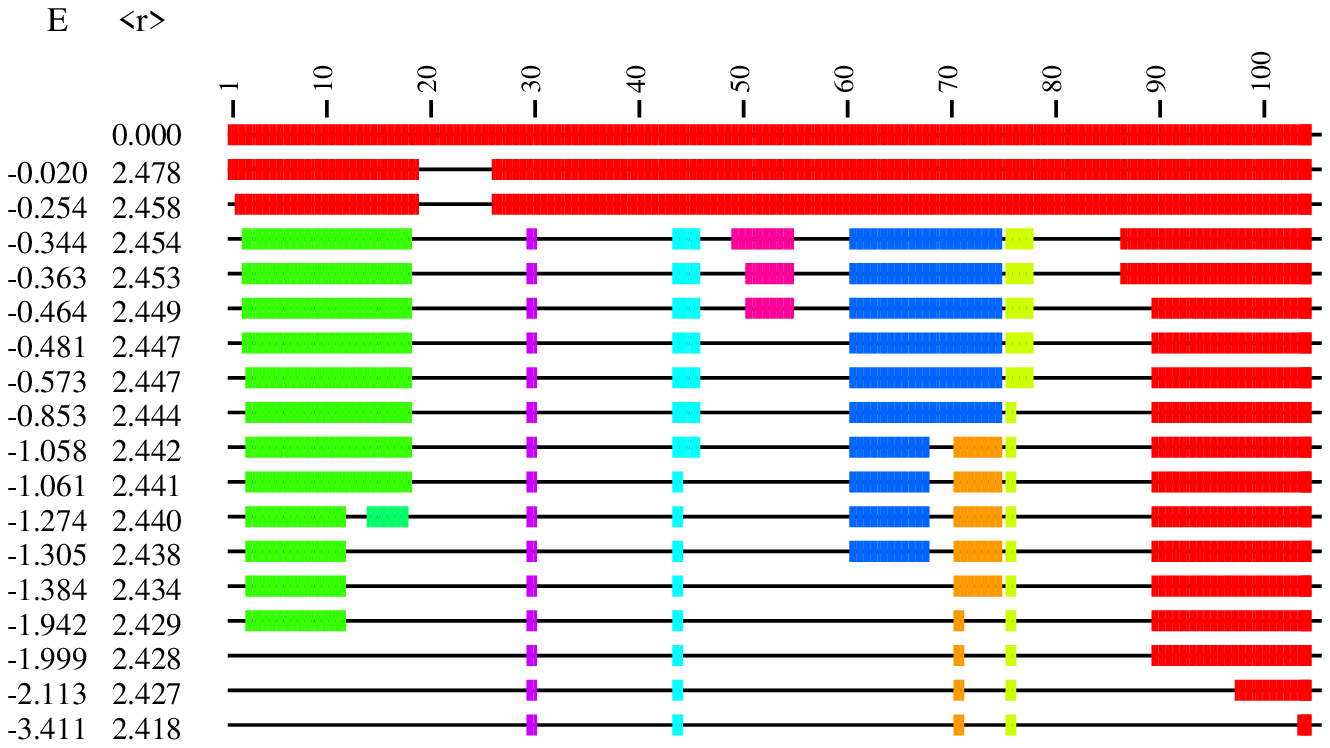} }
\caption{RCD plots for four forms of horse cytochrome C: native, complexed with antibody, complexed with peroxidase, low ionic strength. The large variations in rigidity patterns arise from relatively small variations in protein crystal structure. \label{fig-horse-stripy}}
\end{center}
\end{figure}

\begin{figure}[tb]
\begin{center}
\includegraphics[width=0.99\textwidth, angle=0]{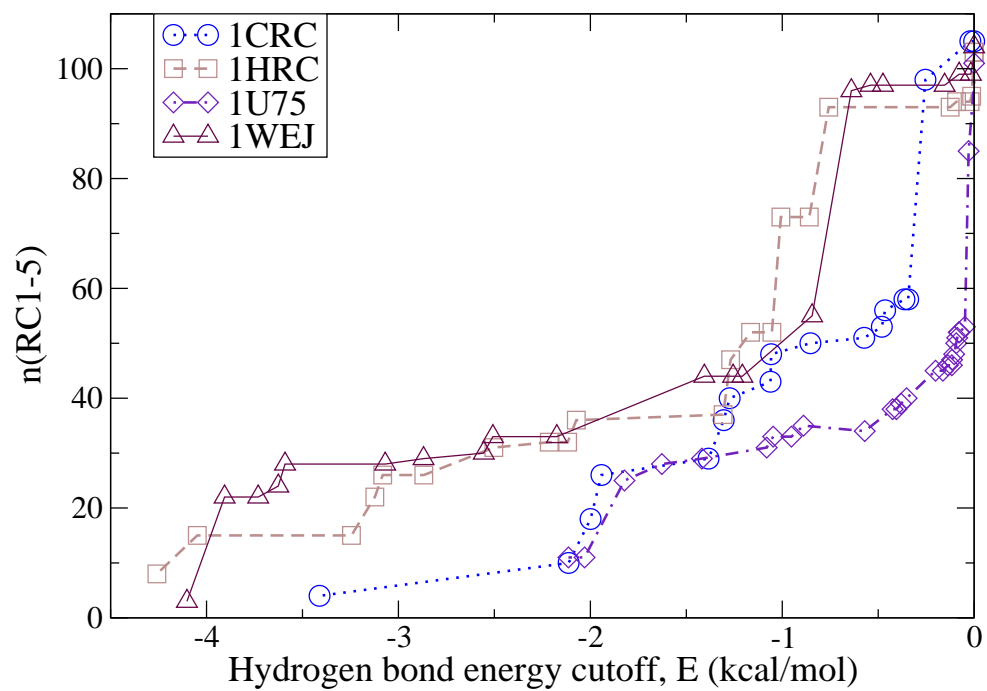}
\caption{ A graph of hydrogen bond energy cutoff $E$ versus rigidity for horse cytochromes. The different symbols and lines denote different types. \label{fig-cyto-en-horse1}}
\end{center}

\end{figure}

In figure \ref{fig-horse-stripy} we show rigidity breakdowns for four forms of horse cytochrome C, namely, the uncomplexed 1HRC structure, the 1WEJ structure which is crystallised in complex with an antibody, the 1U75 structure from a complex with cytochrome C peroxidase, and the low-ionic-strength 1CRC form. The overall rigidity of the four structures is shown in figure \ref{fig-cyto-en-horse1}. Significant differences are visible in the rigidity dilution behaviour of the four structures. For example, the 1HRC and 1WEJ structures retain a significant amount of rigidity (about 25\% of alpha carbons lying within large rigid clusters) at cutoff values as low as around -3 kcal/mol, where the 1U75 and 1CRC structures have lost almost all rigidity.

Let us now investigate the structural variations which give rise to this variation in rigidity behaviour. We can quantify the differences between the structures by aligning the alpha-carbon atoms defining the protein backbone and obtaining the root-mean-square deviation (RMSD) between alpha-carbon positions. We carry this out using the PyMOL ``align'' command \cite{Del____}. The results are given in table \ref{tab-horse-1}. The variations are remarkably small, the largest being 0.572 \AA {} between 1U75 and 1WEJ. It thus appears that quite large variations in rigidity-dilution behaviour can arise from relatively modest structural variations.

\begin{table}
\caption{\label{tab-horse-1} RMSD variations in alpha-carbon positions among four horse cytochrome C structures, in \AA.}
\begin{indented}
\lineup
\item[]\begin{tabular}{ c c c c }
\br
From$\backslash$To:&1HRC&1CRC&1WEJ\cr
1CRC&0.32&---&---\cr
1WEJ&0.318&0.321&---\cr
1U75&0.472&0.53&0.572\cr
\br
\end{tabular}
\end{indented}
\end{table}

\begin{figure}

\begin{center}
  \subfigure[5CYT: normal ferricytochrome]{\includegraphics[scale=0.44]{\figdir/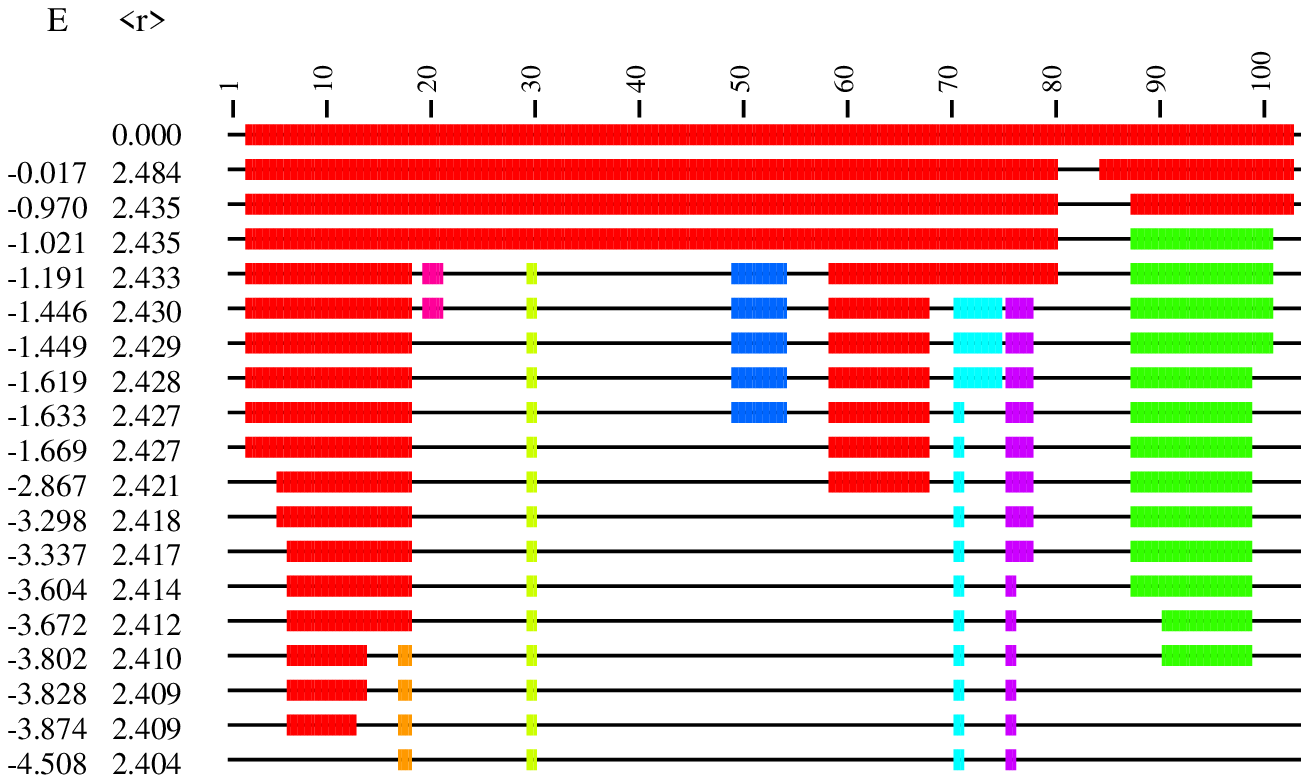} }
  \subfigure[1I55: crystallised from 2Zn:1Fe mix]{\includegraphics[scale=0.44]{\figdir/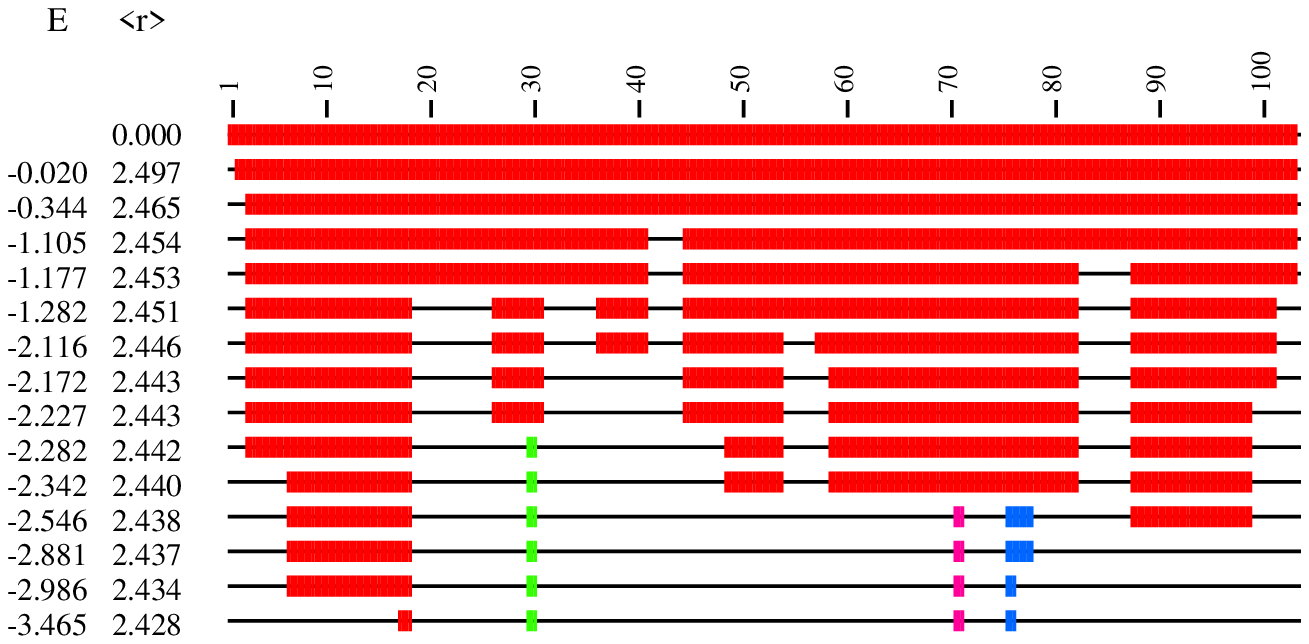} }
  \subfigure[1I54: crystallised from 1Zn:2Fe mix]{\includegraphics[scale=0.44]{\figdir/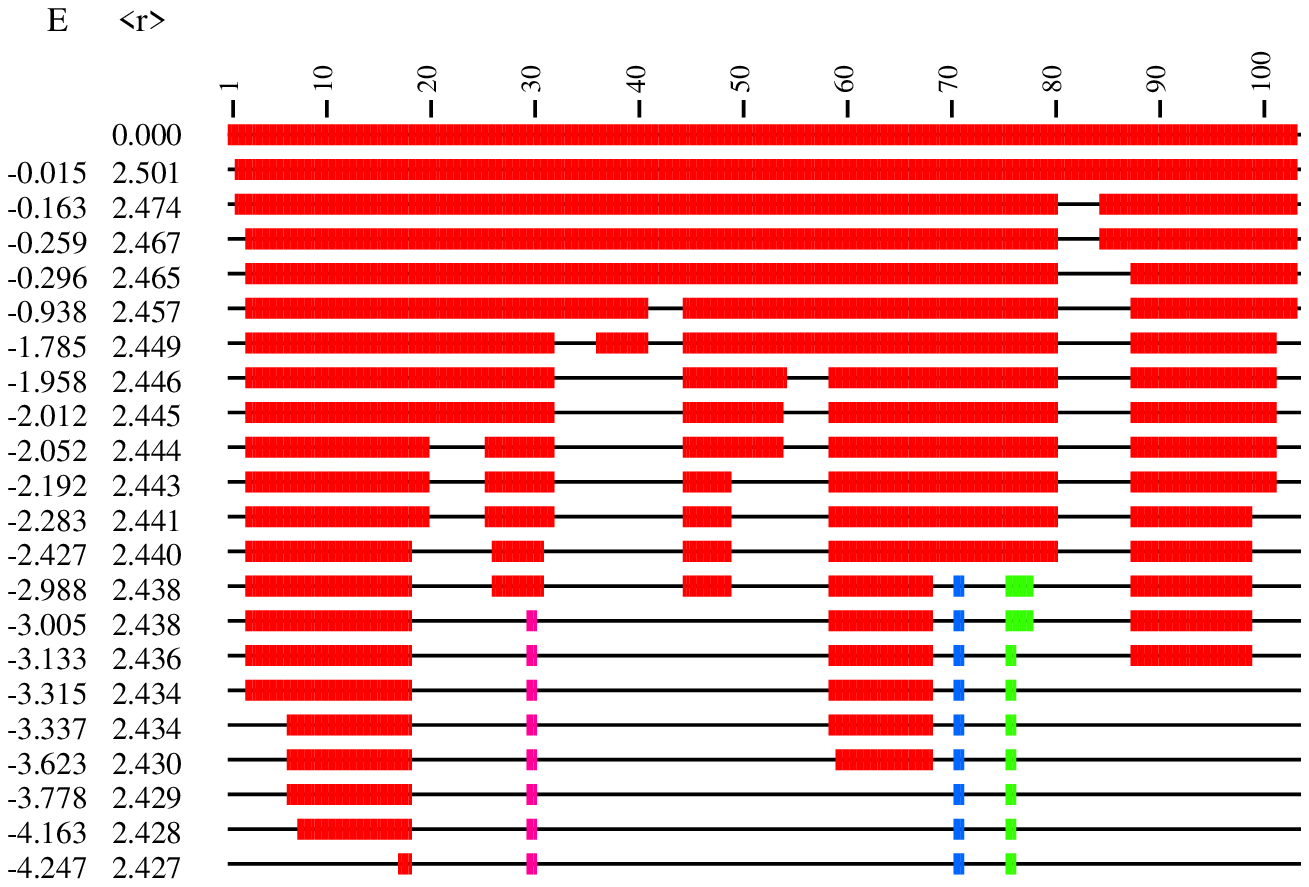} }
  \subfigure[1LFM: with Co replacing Fe]{\includegraphics[scale=0.44]{\figdir/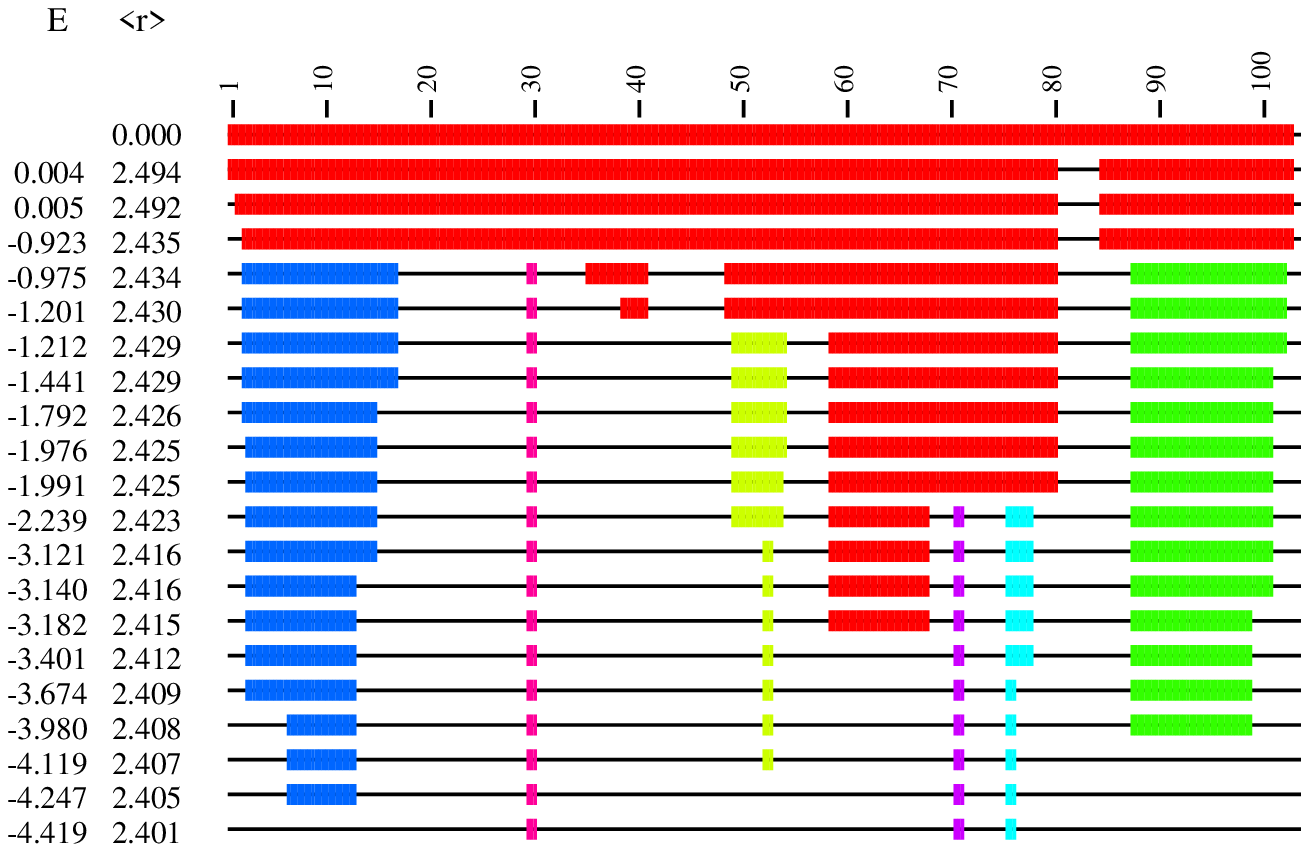} }
\caption{Rigidity dilutions for four forms of tuna cytochrome C crystallised with different metal ion content in the heme groups. (a) normal Fe, (b) from a mixture with 2Zn:1Fe, (c) from a mixture with 2Fe:1Zn, (d) with Co. \label{fig-tuna-stripy}}
\end{center}
\end{figure}

\begin{figure}
\begin{center}
\includegraphics[width=0.99\textwidth, angle=0]{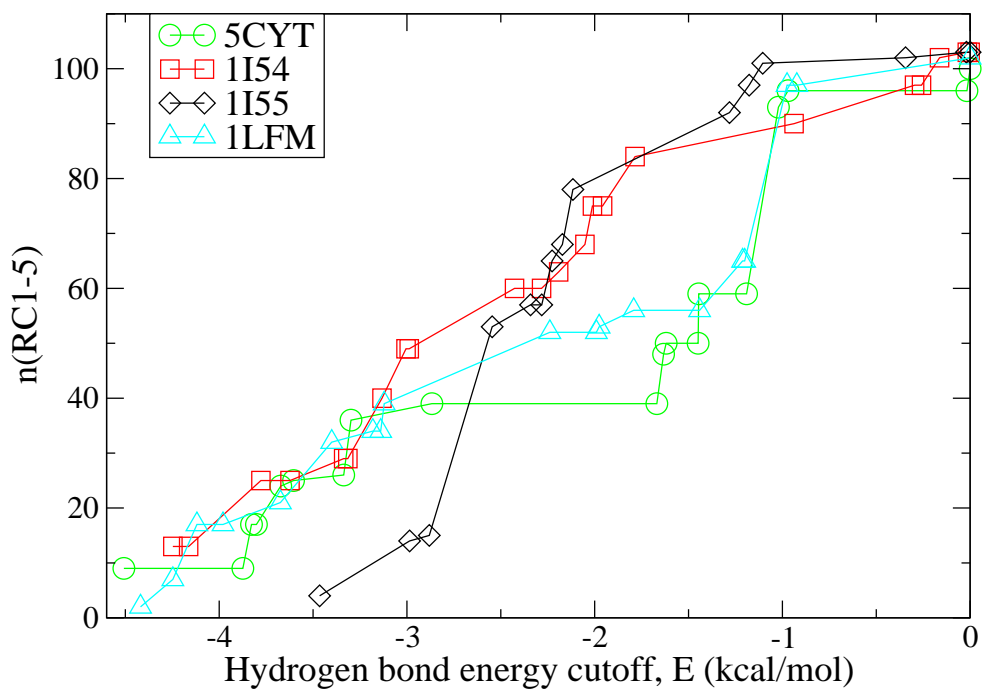}
\caption{ A graph of hydrogen bond energy cutoff $E$ versus rigidity for four tuna cytochrome C structures. Note the considerable differences in behaviour between, for example, 5CYT and 1I55, especially in the $-1$ to $-2$ kcal/mol energy range. \label{fig-cyto-en-tuna1}}
\end{center}

\end{figure}

In figure \ref{fig-tuna-stripy} we show RCD plots for the rigidity dilutions of four mitochondrial cytochrome C structures obtained from tuna, chosen as they differ in their heme-group metal content. 5CYT is a normal ferrocytochrome C with an Fe atom in the heme group. 1I55 and 1I54 were crystallised from mixed-metal cytochromes, with 2Zn:1Fe and 2Fe:1Zn respectively. Evidently a given heme group can contain only one metal atom --- from the pdb files we selected chains containing Fe-heme groups, to maximise comparability with 5CYT. 1LFM, meanwhile, has Co replacing Fe in the heme group. It is clear from comparison of the plots in figure \ref{fig-tuna-stripy} and the corresponding graph of rigidity versus cutoff energy (Figure \ref{fig-cyto-en-tuna1}) that there are noticeable differences between the rigidity behaviour of these four structures. It is particularly instructive to compare the behaviour of 5CYT against that of 1I55 in figure \ref{fig-cyto-en-tuna1}; in the energy range of -1 to -2 kcal/mol, 1I55 remains largely rigid while 5CYT has lost a considerable amount of rigidity. The alpha-carbon RMSD differences between the structures are given in table \ref{tab-tuna-1} and are even smaller than those among the horse cytochrome structures, none being higher than 0.3 \AA.

\begin{table}
\caption{\label{tab-tuna-1} RMSD variations in alpha-carbon positions among four tuna cytochrome C structures, in \AA.}
\begin{indented}
\lineup
\item[]\begin{tabular}{ c c c c }
\br
From$\backslash$To:&5CYT&1I55&1I54\cr
1I55&0.27&---&---\cr
1I54&0.2668&0.041&---\cr
1LFM&0.286&0.116&0.087\cr
\br
\end{tabular}
\end{indented}
\end{table}

\begin{figure}
\begin{center}
\includegraphics[width=0.99\textwidth, angle=0]{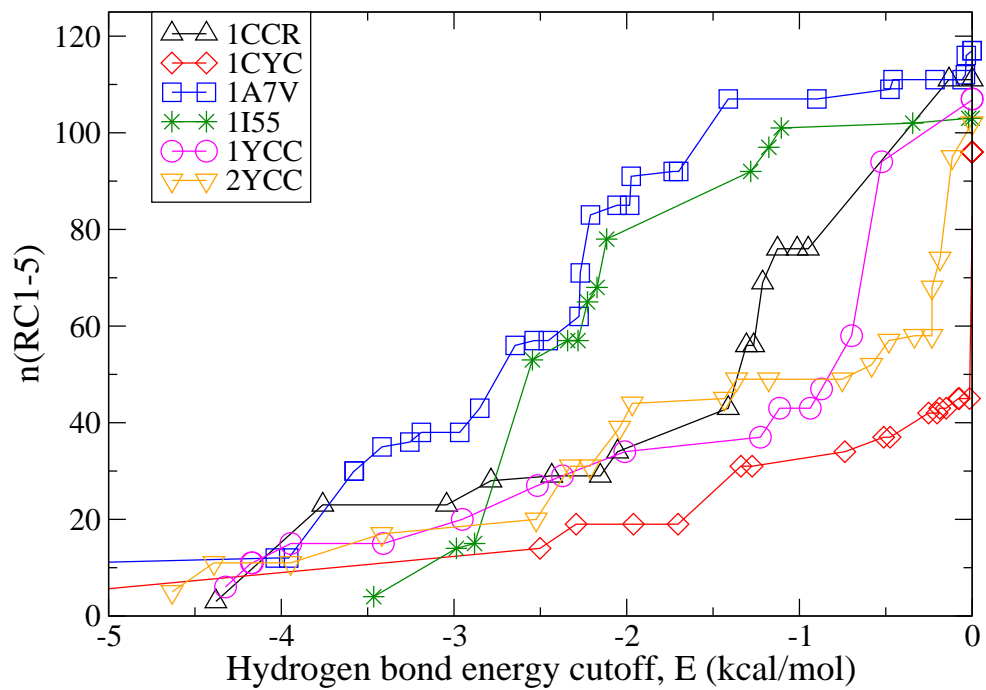}
\caption{ A graph of hydrogen bond energy cutoff $E$ versus rigidity for various cytochromes: 1CCR, rice; 1CYC, bonito; 1A7V, bacterial; 1I55, tuna; 1YCC and 2YCC, yeast. Note the wide variety in rigidity behaviour. \label{fig-cyto-en-other}}
\end{center}

\end{figure}

In figure \ref{fig-cyto-en-other} we show the rigidity versus cutoff energy behaviour of a selection of cytochrome C structures from several different branches of the tree of life. The structures 1I55 (tuna) and 1CYC (bonito) are animal; 1YCC and 2YCC (yeast) are fungal; 1CCR (rice), a plant; and 1A7V, a bacterial cytochrome C structure. It appears that the range of behaviours spans the gamut of possibilities, with some structures rapidly losing rigidity at cutoffs around $-0.5$ kcal/mol and other retaining a large amount of rigidity at cutoffs below $-2$ kcal/mol. In terms of retention of rigidity the 1A7V bacterial structure and the 1I55 tuna structure appear similar, whereas by structure the bacterial protein is quite dissimilar to any of the eukaryotic mitochondrial cytochrome C structures.

Our initial hypothesis of a correlation between rigidity behaviour and protein function thus appears to have been disconfirmed by the cytochrome C data. Instead, protein rigidity appears to depend on quite small structural variations; it varies widely within a family of closely related, isostructural and isofunctional proteins.

\subsection{Hemoglobin, myoglobin and lactalbumin}
\label{sec-multi}

\begin{figure}
\begin{center}
\includegraphics[width=0.7\textwidth, angle=270]{\figdir/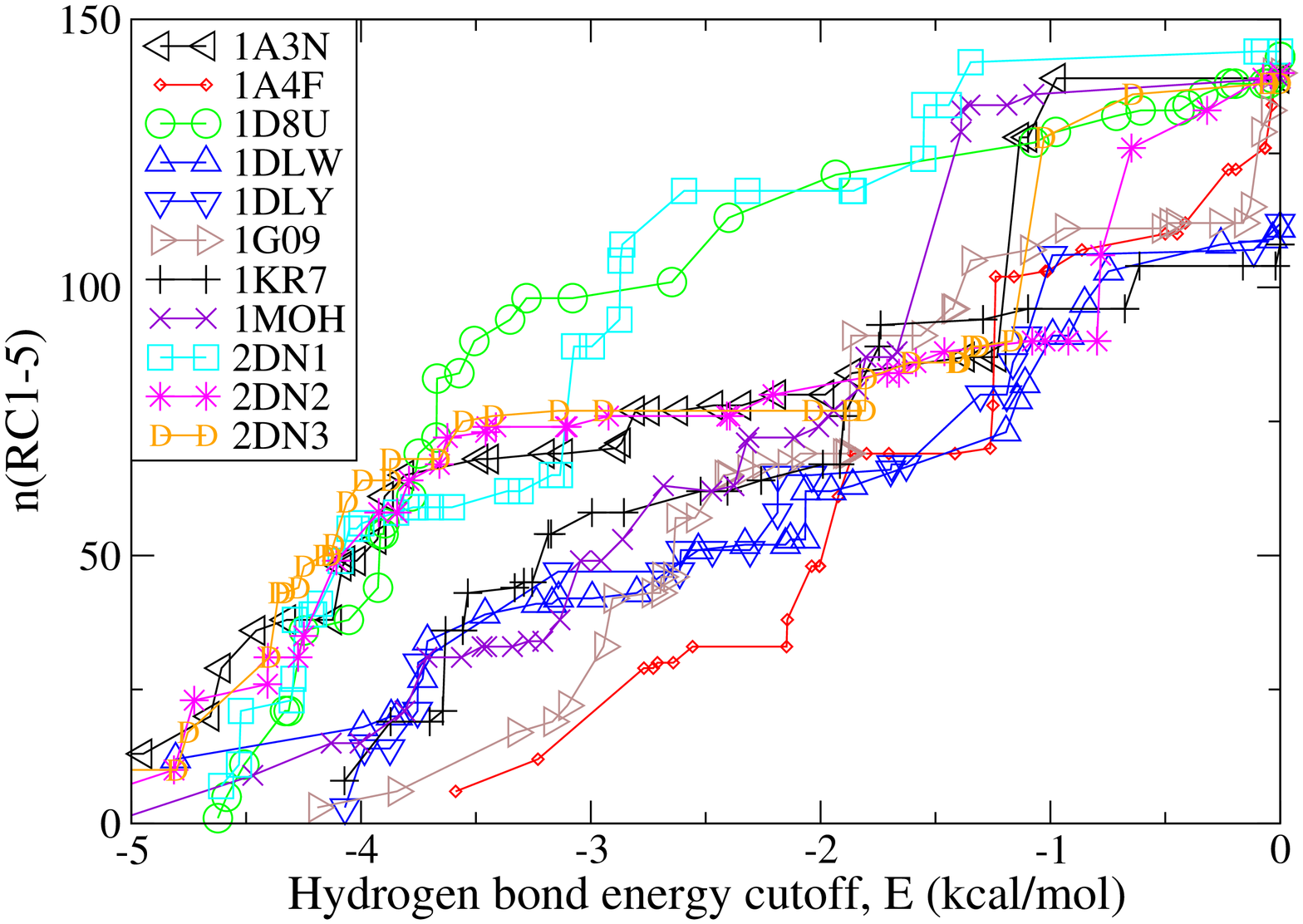}
\caption{ A graph of hydrogen bond energy cutoff $E$ versus rigidity for a variety of hemoglobins, showing considerable variation in the pattern of rigidity loss during dilution. 1A3N, 2DN1-alpha, 2DN2 and 2DNS are all human hemoglobin alpha subunits; 2DN2-beta is a beta subunit.  \label{fig-hemo-en}}
\end{center}
\end{figure}

\begin{figure}
\begin{center}
\includegraphics[width=0.99\textwidth, angle=0]{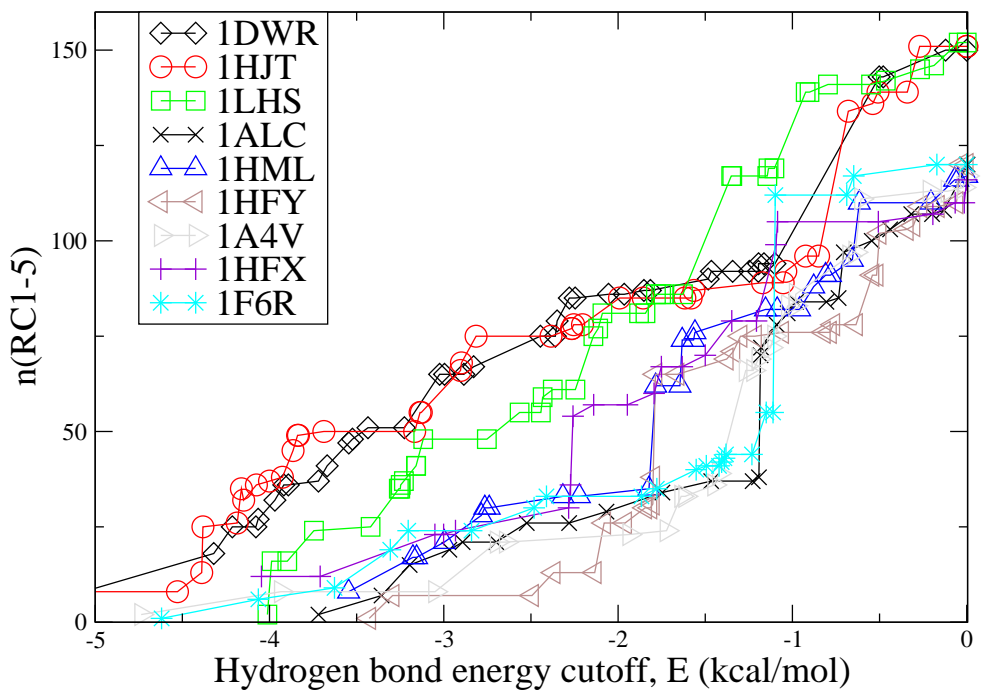}
\caption{ A graph of hydrogen bond energy cutoff $E$ versus rigidity for a variety of myoglobins and alpha-lactalbumins, showing considerable variation in the pattern of rigidity loss during dilution. \label{fig-multi-en}}
\end{center}

\end{figure}

\begin{table}
\caption{\label{tab-hemo-1} RMSD variations in alpha-carbon positions among four human alpha-hemoglobin structures, in \AA.}
\begin{indented}
\lineup
\item[]\begin{tabular}{ c c c c }
\br
From$\backslash$To:&2DN1&2DN2&2DN3\cr
1A3N&0.445&0.219&0.446\cr
2DN1&---&0.527&0.215\cr
2DN2&---&---&0.479\cr
\br
\end{tabular}
\end{indented}
\end{table}

In figure \ref{fig-hemo-en} we show rigidity versus cutoff energy $E$ for 11 hemoglobin structures displaying a wide variation in rigidity behaviour, while in figure \ref{fig-multi-en} we show results for a set of three myoglobin structures and six alpha-lactalbumins from a variety of mammals.
In each case we see wide variation in the pattern of rigidity loss during the dilution, with some members of each family largely losing rigidity by the time the cutoff energy $E$ drops to around -2 kcal/mol and with others retaining considerable rigidity down to much lower cutoff values.

For the hemoglobins, we note that our dataset includes four human hemoglobin structures. Those identified in Figure \ref{fig-hemo-en} as 1A3N, 2DN1-alpha, 2DN2 and 2DN3 are alpha subunits of human deoxy, oxy, deoxy and carbonmonoxy-hemoglobin respectively. Interestingly, their rigidities differ for cutoff values in the range 0 to -2 kcal/mol but are all similar for lower cutoff values, suggesting that their structural differences (see table \ref{tab-hemo-1} for alpha-carbon RMSD values) mainly affect the weaker hydrogen bonds. For comparison, we also show data for 2DN1-beta, a beta subunit from the 2DN1 structure, whose rigidity clearly differs from any of the human alpha subunits, resembling that of the unrelated rice hemoglobin structure 1D8U.

\subsection{Trypsin}
\label{sec-trypsin}

\begin{figure}
\begin{center}
\includegraphics[width=0.99\textwidth, angle=0]{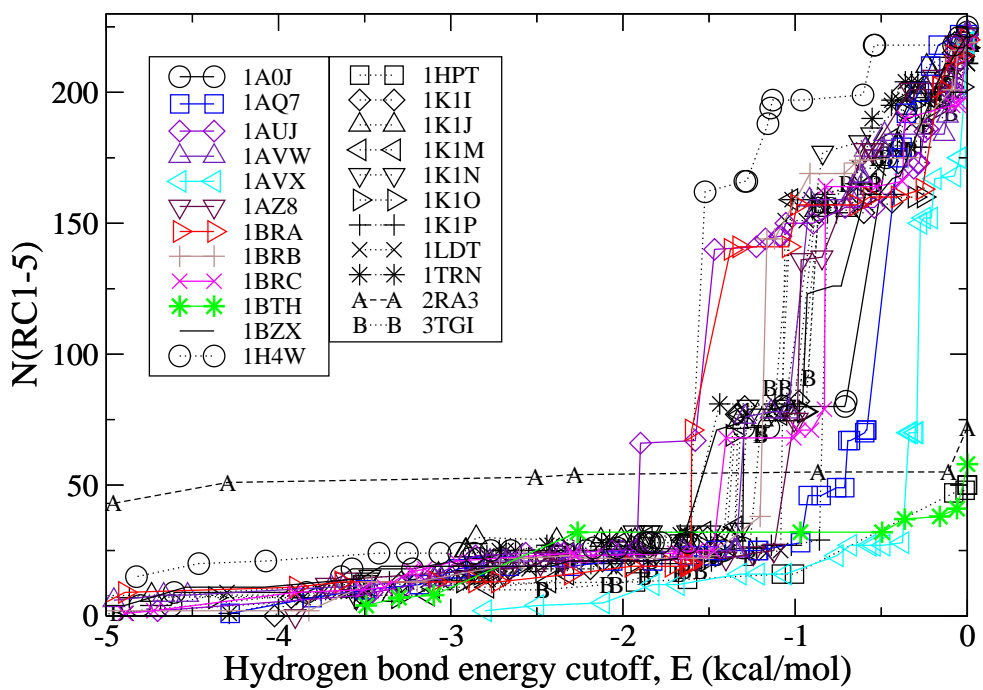}
\caption{A graph of hydrogen bond energy cutoff $E$ versus rigidity for a wide variety of trypsin structures. Notice, in contrast to previous figures, that all trypsins lose rigidity for cutoffs in the region of -1 to -2 kcal/mol. \label{fig-trypsin}}
\end{center}
\end{figure}

The presence of a trypsin inhibitor, BPTI, in the Hespenheide dataset \cite{HesRTK02} suggested to us that we should include trypsin itself in our study. Many crystal structures of trypsins have been reported from different species and under multiple conditions of crystallisation (e.g.\ in complex with BPTI or various small molecules); we assembled a dataset of 18 structures from five different species- human, bovine, rat, pig and salmon.

The trypsins display a unique feature in their rigidity-dilution behaviour which sets them aside from any of our other proteins, as shown in Figure \ref{fig-trypsin}. They display a steep loss of rigidity at relatively low values of the hydrogen-bond energy cutoff, in the range of -1 to -2 kcal/mol. Comparison to previous figures shows clearly how much this behaviour differs from the other protein groups we have examined.

Trypsin is the largest of the proteins in our set, typical structures having about 220 residues. This suggests that the structure might divide into a larger number of medium-sized rigid clusters during the dilution, causing us to `lose' some atoms when we count only the five largest rigid clusters. Inspection of the RCD plots, however, shows that trypsin does not appear very different to other proteins in our set and does not divide into a larger number of sizeable rigid clusters. It thus appears that there is genuinely something different about the behaviour of the trypsin fold during rigidity dilution.

The distinctiveness of the trypsins is only evident because of our strategy of comparing multiple related structures from different organisms and conditions. A single trypsin structure might not have appeared unusual when compared against single examples of other proteins. Our comparative approach, however, shows clearly that our other protein families display much more variation in the rigidity dilution behaviours and generally include some members which retain rigidity down to much lower energy cutoffs, e.g.\ $-3$ to $-5$ kcal/mol. This result thus confirms the value of the present comparative approach.

\section{Conclusion and outlook}
\label{sec-concl}

Our motivation in this study was twofold. Firstly, we wished to clarify a methodological issue in the use of rigidity analysis on protein structures, by determining the robustness of RCDs against small structural variations. Secondly, we wished to obtain, from our large sample of protein structures, an insight into the ``typical'' pattern of rigidity loss during hydrogen-bond dilution.

On the first point, we find that there is considerable variation in the RCDs of structurally similar proteins during dilution. Figure \ref{fig-cyto-en-other}, for example, shows that among a group of cytochrome C structures drawn from various eukaryotic mitochondria, energy cutoffs in the range from 0 to $-2$ kcal/mol (such as have typically been used for FIRST/FRODA simulations of flexible motion \cite{WelMHT05, JolWHT06, JolWFT08, MacNBC07}) can produce anything from near-total rigidity to near-total flexibility. We conclude that the results of rigidity analysis on individual crystal structures should not be over-interpreted as being ``the'' RCD for a protein.

This implies that rigidity analysis is best used for comparison of structures and for hypothesis testing. {\it Comparison} of the RCDs of two similar structures can bring out the significance of structural variation, by drawing attention to those parts of the protein where the constraint network has changed. In combination with simulation of flexible motion, a {\it hypothesis testing} approach can reveal whether the presence or absence of a particular interaction is significant in allowing or forbidding a flexible motion to occur. An early use of this approach was the study by Jolley et al.\ \cite{JolWHT06}, which compared two hypothetical pathways for formation of a protein complex by examining the changes in rigidity and flexibility which each pathway would entail.

On the second point, we found that for most of the protein structures in our survey, rigidity loss occurs gradually and continuously as a function of energy cutoff during dilution. The only case which stood out was that of the enzyme trypsin. We surveyed a large number of trypsin structures from multiple organisms and found that all of the structures displayed near-total loss of rigidity before the cutoff dropped to $-2$ kcal/mol. No other protein group in our survey displayed such behaviour. We cannot at present account for this distinctive phenomenon, which may imply a difference in the structure-function relationship in the trypsins compared to other proteins.

Our results in this paper suggest two avenues for further enquiry. Firstly, the rigidity of protein monomers extracted from complexes should be compared with their rigidity within the complex, which will be affected by interactions between monomers. Secondly, the robustness of flexible motion simulations based on rigidity analysis must now be investigated.
A recent study of the flexible motion of myosin \cite{SunRAJ08} found that the flexible motion of the myosin structure appeared qualitatively similar over a wide range of cutoff values covering both highly flexible and highly rigid structures. This suggests that rigidity analysis retains its value as a natural coarse-graining for simulations even if the rigidity behaviour during dilution is as variable as we have found.



\ack
We thankfully acknowledge discussions with R.\ Freedman and T.\ Pinheiro. SAW and RAR gratefully acknowledge the Leverhulme Trust (grant F/00 215/AH) for financial support.

%


\end{document}